\documentclass[journal, twoside, balance]{IEEEtran}
\usepackage{cite}

\RequirePackage{amsthm}

\IEEEoverridecommandlockouts
\usepackage{tablefootnote}
\usepackage{bm}
\usepackage{algorithm, algorithmic}
\usepackage{cite}
\usepackage{amsmath,amssymb,amsfonts}
\usepackage{algorithmic}
\usepackage{graphicx}
\usepackage{textcomp}
\usepackage{xcolor}
\usepackage{setspace}
\usepackage{soul} 
\usepackage{color, xcolor} 
\usepackage{graphicx}
\usepackage{epstopdf}
\usepackage{pdfpages}
\usepackage{tabularx}

\usepackage{graphicx}
\usepackage{float}
\usepackage{subfig}

\usepackage{grffile}
\usepackage{enumitem}

\usepackage{array}    
\usepackage{amsmath}  
\usepackage{multirow}

\usepackage{booktabs}

\usepackage{makecell}
\usepackage{comment}

\usepackage{dblfloatfix}

\usepackage{multicol} 

\usepackage{balance}

\def\BibTeX{{\rm B\kern-.05em{\sc i\kern-.025em b}\kern-.08em
 T\kern-.1667em\lower.7ex\hbox{E}\kern-.125emX}}

\newtheorem{theorem}{Theorem}
\newtheorem{proposition}{Proposition}
\newtheorem{lemma}{Lemma}

\theoremstyle{definition}

\theoremstyle{remark}

\begin{document}

\title{Precise Analysis of Covariance Identifiability for Activity Detection in Grant-Free Random Access}

\author{ Shengsong~Luo,
         Junjie~Ma, \textit{Member, IEEE},
         Chongbin~Xu, \textit{Member, IEEE},
		Xin~Wang, \textit{Fellow, IEEE} 
  \thanks{

S. Luo, C. Xu, and X. Wang are with the Key Lab of EMW Information (MoE), the Department of Communication Science and Engineering, Fudan University, Shanghai 200433, China (e-mails: \{22110720120\}@m.fudan.edu.cn, \{chbinxu, xwang11\}@fudan.edu.cn). 

J. Ma is with the State Key Laboratory of Scientific and Engineering Computing, Institute of Computational Mathematics and Scientific/Engineering Computing, Academy of Mathematics and
 Systems Science, Chinese Academy of Sciences, Beijing 100190, China (e-mail: majunjie@lsec.cc.ac.cn).
}

}

\maketitle

\begin{abstract} 
We consider the identifiability issue of maximum-likelihood based activity detection in massive MIMO-based grant-free random access. An intriguing observation by Chen \textit{et al.} \cite{Phase_Yuwei_2022} indicates that the identifiability undergoes a phase transition for commonly-used random user signatures as $L^2$, $N$ and $K$ tend to infinity with fixed ratios, where $L$, $N$ and $K$ denote the user signature length, the total number of users, and the number of active users, respectively. In this letter, we provide a precise analytical characterization of the phase transition based on a spectral universality conjecture. Numerical results demonstrate excellent agreement between our theoretical predictions and the empirical phase transitions.
\end{abstract}


\section{Introduction}\label{Introduction}

Massive machine-type communications (mMTC) involve a large number of sporadically active devices \cite{mMTC2016, mMTC2018, mMTC_mahmood2020white, Chen2021Massive}. In this letter, we focus on the uplink transmission of mMTC and study the \textit{activity detection} problem where the base station (BS), equipped with a large number of antennas, aims to identify the active devices among a bulk of potential users.


Assuming independently and identically distributed (IID) fast fading channel coefficients, the received signals at different antennas of the BS can be viewed as {independent} samples drawn from a common zero-mean Gaussian distribution, whose covariance matrix depends on the unknown activity pattern and the user signatures. Under such an assumption, the sample covariance matrix is a sufficient statistic for the activity pattern \cite[Theorem 1]{MLE1_sufficient}. For this reason, the maximum likelihood estimator (MLE) is also referred to as the \textit{covariance-based approach} \cite{MLE1_sufficient, Non_Bayesian_2021, Khanna_supportRecovery2022} in the literature of grant-free random access. 

The large sample limit of the MLE provides important insight for its performance under massive MIMO. (Here the large sample limit corresponds to the infinite-antenna limit). It is well-known that the MLE converges (in probability) to the true parameters, provided that certain regularity conditions hold \cite{kay1993fundamentals}. Among these conditions, the main requirement is that the model must be \textit{identifiable}, namely, the map from the parameters to the distribution is ``injective''. For the non-negative MLE formulation considered in \cite{Non_Bayesian_2021}, the {identifiability} of the model eventually reduces to the injectivity of a linear map under non-negativity constraint \cite{Phase_Yuwei_2022}, which can be verified numerically using linear programming (LP) \cite{Phase_Yuwei_2022}. This problem has also been studied in the array signal processing literature \cite{Piya_correlation_aware2013, Piya_limit2018}. 

Intriguingly, Chen \textit{et al.} \cite{Phase_Yuwei_2022} observed that the identifiability event undergoes a phase transition for commonly-used random user signatures, as $L^2$, $N$ and $K$ tend to infinity with fixed ratios, where $L$, $N$ and $K$ denote the signature length, the number of total users, and the number of active users, respectively. Note that the linear map involved in the identifiablity event depends on the user signatures through a complicated column-wise Kronecker product (Khatri–Rao product) structure, which poses a significant challenge for rigorous analysis of the phase transition phenomenon. In this letter, inspired by \cite[Section 4.2]{donoho2010counting} and the recent spectral universality principle established in \cite{Rishabh2024_Spectral_Universality}, we introduce a surrogate model (coined ``semi-random" model) for the Khatri-Rao product model, which faithfully captures the structure of the linear map and becomes mathematically tractable. Using tools developed in \cite{Tropp}, we provide a precise analytical characterization of the phase transition phenomenon for the semi-random model. Numerical results suggest that our theoretical predictions accurately describe the empirical phase transitions for various random user signatures.


\section{Problem Description}


Consider an uplink transmission from a pool of $N$ single-antenna users to an $M$-antenna BS. Assume that only $K$ of the $N$ users are active. Each user is uniquely determined by the signature $\boldsymbol{s}_n \in \mathbb R^{L}$, $1\leq n \leq N$. Let $u_n \in \{0,1\}$ be an indicator variable for the activity of user $n$, namely, $u_n=1$ if user $n$ is active, and $u_n = 0$ otherwise. 
We model a Rayleigh fading channel experienced by the $m^{\text{th}}$ antenna as $h_{n,m} = \beta_n {g}_{n,m}$ for user $n$, where ${g}_{n,m} \sim \mathcal{N}({0}, 1)$ and $\beta_n$ represents the large-scale fading component.
The received signal at the $m^{\text{th}}$ antenna is
\begin{equation} \label{receivesignal}
\begin{aligned}
\boldsymbol{y}_m &= \boldsymbol{S} \text{diag}(\boldsymbol{\gamma}^{\frac{1}{2}}) \boldsymbol{h}_m + \boldsymbol{w}_m, \ m=1,\ldots,M,
\end{aligned}
\end{equation}
where $\boldsymbol{S} = [\boldsymbol{s}_1, \cdots, \boldsymbol{s}_N] \in \mathbb R^{L \times N}$ is the signature matrix, $\bm h_m = [h_{1,m}, \cdots, h_{N,m}]^T \in \mathbb R^{N}$ with $\bm h_m \sim \mathcal{N}(\bm 0, \bm I)$, 
$\text{diag}(\boldsymbol{\gamma}^{\frac{1}{2}})$ is a diagonal matrix whose diagonal elements are the entry-wise square root of $\boldsymbol{\gamma}=[\gamma_1, \cdots, \gamma_N]^T \in \mathbb R_{+}^{N}$ with $\gamma_n=(u_n \beta_n)^2$, and $\bm w_m \sim \mathcal{N}(\bm 0, \sigma^2 \bm I)$. 

{Activity detection} refers to the detection of the index set of the nonzero components in $\boldsymbol \gamma$. Following \cite{Non_Bayesian_2021}, we consider a MLE framework, where $\boldsymbol{\gamma}$ is treated as a deterministic yet unknown parameter. Given $\bm{\gamma}$, the received signals $\{\bm y_m\}_{m=1}^M$ at all antennas are IID, following a common Gaussian distribution $\mathrm{Pr}(\bm{y};\bm{\gamma}):= \mathcal{N}(\boldsymbol{0}, \boldsymbol{\Sigma}_\gamma)$, where $\boldsymbol{\Sigma}_\gamma=\boldsymbol{S} \text{diag}(\boldsymbol{\gamma}) \boldsymbol{S}^T + \sigma^2 \boldsymbol{I}$.

In this letter, we focus on the {identifiability} issue of the MLE. Let $\boldsymbol \gamma^{\circ}\in\mathbb{R}_+^N$ be the true parameter which satisfies $\Vert \boldsymbol \gamma^{\circ} \Vert_0 = K$. For computational tractability, we consider a constrained MLE formulation that takes into account the non-negativity of the parameter $\bm{\gamma}$, but not its sparsity, i.e.,

\begin{equation}
    \max_{\bm \gamma \in \mathbb R^N_{+}} \ \prod_{m} \mathrm{Pr} \left( \bm y_m; \bm \gamma \right).
\end{equation}

Since the likelihood function $\mathrm{Pr}(\boldsymbol{y}; \boldsymbol \gamma)$ is a Gaussian distribution function with covariance $\boldsymbol{\Sigma}_\gamma=\boldsymbol{S} \text{diag}(\boldsymbol{\gamma}) \boldsymbol{S}^T + \sigma^2 \boldsymbol{I}$, 
the {identifiablity} (at the ground truth $\bm{\gamma}^\circ$) reduces to the following covariance identifiablity \cite{Phase_Yuwei_2022}:
\begin{equation} \label{identifiablity_equation}
    \nexists \gamma \in \mathbb R^{N}_{+} \setminus \{\boldsymbol \gamma^{\circ}\} \Longrightarrow  \boldsymbol{S} \text{diag}(\boldsymbol{\gamma}^{\circ}) \boldsymbol{S}^T = \boldsymbol{S} \text{diag}(\boldsymbol{\gamma}) \boldsymbol{S}^T.
\end{equation}
This can be equivalently written as follows \cite{Phase_Yuwei_2022}
\begin{equation} \label{convex_feasibility_event}
\begin{aligned}
   \mathcal{N}(\boldsymbol{A})  \cap \mathcal{C} = \{\boldsymbol{0}\},
\end{aligned}
\end{equation}
with
\begin{equation}
    \begin{aligned}
        \boldsymbol{A} & := [ \boldsymbol{s}_1 {\otimes} \boldsymbol{s}_1, \cdots, \boldsymbol{s}_N {\otimes} \boldsymbol{s}_N ] \in \mathbb R^{L^2 \times N}, \\
        \mathcal{C} & := \{ \boldsymbol x \in \mathbb R^N \mid \boldsymbol{x}_{\mathcal{I}} {\ge} \boldsymbol{0}_{N-K} \},
    \end{aligned}
\end{equation}
where $\mathcal{N}(\bm A)$ is the null space of $\bm A$, $\otimes$ is the Kronecker product, and $\mathcal{I}$ is the index set for the zero entries of $\boldsymbol{\gamma}$. The cone $\mathcal{C}$ is the set of feasible directions at $\bm{\gamma}^\circ$. 
Let $d:=\frac{L^2-L}{2}$. Specifically, 
\begin{equation} \label{compact_null}
    \mathcal{N}(\boldsymbol{A})= \left\{ \boldsymbol{x}\in \mathbb{R} ^N \vline \left( \begin{array}{c}
	\boldsymbol{A}_1\\
	\boldsymbol{A}_2\\
\end{array} \right) \boldsymbol{x}=\boldsymbol{0} \right\} ,
\end{equation}
in which $\boldsymbol{A}_1 := \boldsymbol{S} \odot \boldsymbol{S}$, $\boldsymbol{A}_2 := [\bm s_{1} \tilde{\otimes} \bm s_1, \cdots, \bm s_{N} \tilde{\otimes} \bm s_N]\in \mathbb R^{d \times N}$. Here $\odot$ is the Hadamard product, $\bm s_{n} \tilde{\otimes} \bm s_n (\forall n)$ denotes a $d$-dimensional vector which contains the lower-triangular elements of the matrix $\bm s_n \bm s_n^T$ (excluding the diagonal). We refer to \cite[Definition 5]{Piya_limit2018} for a similar representation.

Our goal is to provide a precise phase transition characterization (in terms of the parameters $L,K,N$) for \eqref{convex_feasibility_event} with randomly generated $\bm{S}$.


\section{Semi-Random Model: Phase Transition Analysis }

In this section, we introduce a \textit{semi-random} model\footnote{It is different from the semi-random matrix model in \cite{dudeja2023universality}.}, which serves as a mathematically tractable model to analyze the phase transition. In Section \ref{Numerical_Results}, we will show that the analytical phase transition for the semi-random model accurately describes empirical results for many commonly-used signature matrices.

\subsection{Semi-Random Model}\label{Sec:semi-random}

To motivate the introduction of the semi-random model, consider a Rademacher model where the elements of $\boldsymbol{S}$ are taken from $\{\pm 1\}$. In this case, we have $\bm{A}_1:=\bm{S}\odot \bm{S}=\bm{1}_{L\times N}$, where $\bm{1}_{L\times N}$ denotes an $L\times N$ all-ones matrix; see \eqref{compact_null}. The null-space of $\bm{A}$ in \eqref{compact_null} is the same as the null-space of $[\bm{1}_{1\times N};\bm{A}_2]$; where $\bm{A}_2$ is defined in \eqref{compact_null}. Motivated by this observation, and inspired by \cite[Section 4.2]{donoho2010counting} and the recent work of \cite{abbara2020universality, dudeja2022universality, dudeja2023universality}, we introduce the following \textit{semi-random} matrix surrogate for $[\bm{1}_{1\times N};\bm{A}_2]$:
\begin{equation} \label{semi_random}
    \boldsymbol{A}_{\text{semi-random}} := \left( \begin{array}{c}
	\boldsymbol{1}_{1\times N}\\
	\boldsymbol{A}_{\text{RI}}
\end{array} \right),
\end{equation}
where \textit{$\boldsymbol{A}_{\text{RI}} \in \mathbb R^{d \times N}$ is a right rotationally-invariant matrix that has the same spectrum as $\bm{A}_2$}. Here, the spectrum refers to the limiting singular value distribution of the matrix. Note that the feasibility problem in \eqref{convex_feasibility_event} is only concerned with the null space.

\textit{Universality Conjecture:} In what follows, we shall focus on the semi-random model in our analysis. Our goal is to establish the precise phase transition condition for \eqref{convex_feasibility_event} using the semi-random model as a surrogate:
\begin{equation} \label{semi_identifiability_event}
    \mathcal N({\boldsymbol{A}_{\text{semi-random}}}) \cap \mathcal{C} = \{ \boldsymbol{0} \}.
\end{equation}

\subsection{Phase Transition Analysis}

Our analysis is based on the following proposition.

\begin{proposition}\label{Pro:Tropp}
\cite[Theorem I]{Tropp}: Let $\bm{V}\in\mathbb{R}^{N\times N}$ be a Haar distributed random orthogonal matrix. Let $\mathcal{D},\mathcal{K}$ be two closed convex cones in $\mathbb{R}^N$. The following holds for any $\eta \in (0,1)$: 
\begin{equation}
    \begin{aligned}
        \frac{1}{N} \left (\delta(\mathcal{D}) +  \delta(\mathcal{K}) \right ) & \leq 1 - \xi_{\eta} \frac{1}{\sqrt{N}} \Rightarrow \mathrm{Pr}\{ \mathcal{D} \cap \boldsymbol{V} \mathcal{ K} \neq \{\boldsymbol{0}\} \} \leq \eta \\
        \frac{1}{N} \left (\delta(\mathcal{D}) +  \delta(\mathcal{K}) \right )  & \geq 1 + \xi_{\eta} \frac{1}{\sqrt{N}} \Rightarrow  \mathrm{Pr}\{ \mathcal{D} \cap \boldsymbol{V} \mathcal{ K} = \{\boldsymbol{0}\} \} \leq \eta,
    \end{aligned}
\end{equation}
where $\xi_{\eta} := \sqrt{8 \log(4/\eta)}$, and $\delta(\mathcal{D})$ (similarly for $\delta(\mathcal{K})$) denotes the statistical dimension of $\mathcal{D}$:
\begin{equation} \label{statistical_dimension_compute}
    \delta(\mathcal{D}) := \mathbb E \left[  \Vert  {\Pi}_{\mathcal{D}} (\boldsymbol{g}) \Vert^2 \right],\  \text{where }\boldsymbol{g}\sim\mathcal{N}(\bm{0},\bm{I}).
\end{equation}
Here, ${\Pi}_{\mathcal{D}}$ denotes the projection onto the cone $\mathcal{D}$. 
\end{proposition}

To apply Proposition \ref{Pro:Tropp} to analyze \eqref{semi_identifiability_event}, we first rewrite \eqref{semi_identifiability_event} as
\begin{equation} 
    \mathcal{N}(\boldsymbol{A}_{\text{RI}}) \cap \mathcal{D} = \{ \boldsymbol{0} \},
\end{equation}
where $\bm A_{\text{RI}}$ is a right rotationally-invariant matrix (see \eqref{semi_random}) and
\begin{equation}\label{Eqn:D_def}
    \mathcal{D} := \mathcal{C} \cap \{ \boldsymbol{x} \mid \boldsymbol{1}^T \boldsymbol{x}=0 \} = \{ 
\boldsymbol{x} \mid \boldsymbol{1}^T \boldsymbol{x}=0, \boldsymbol{x}_{\mathcal{I}} \ge \boldsymbol{0}_{N-K} \}.
\end{equation}

\begin{figure*}[htbp]
	\centering
	\subfloat[{Gaussian}]{\includegraphics[width=0.28\textwidth]{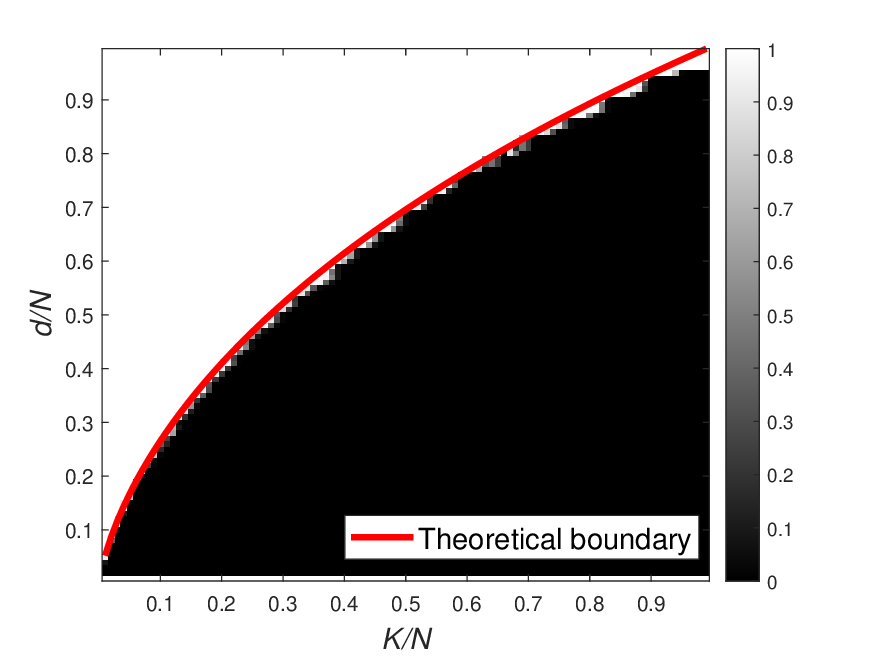}}
	\hfil
	\subfloat[{Rademacher}]{\includegraphics[width=0.28\textwidth]{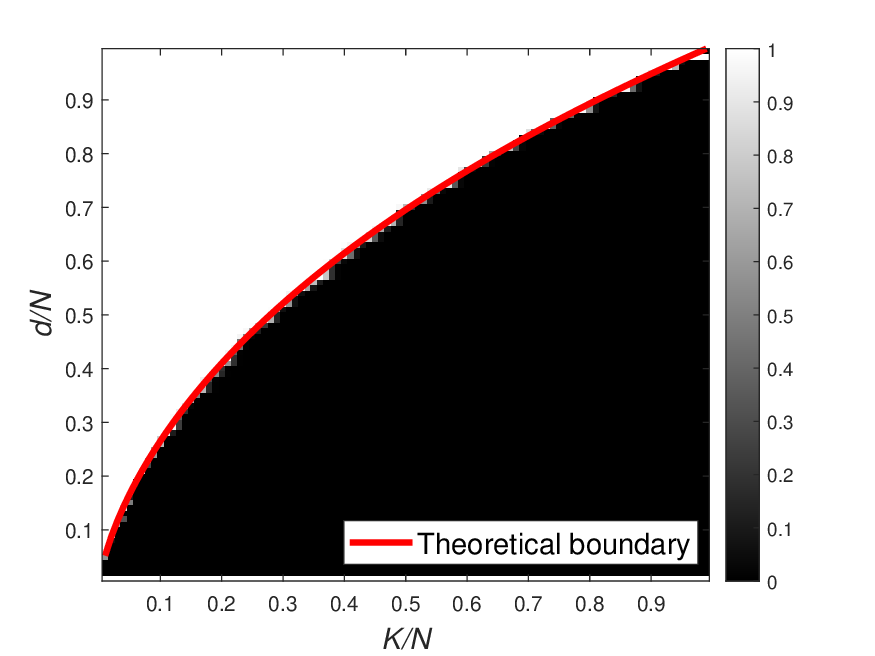}}
	\hfil
	\subfloat[{Doubly sub-sampled Hadamard}]{\includegraphics[width=0.28\textwidth]{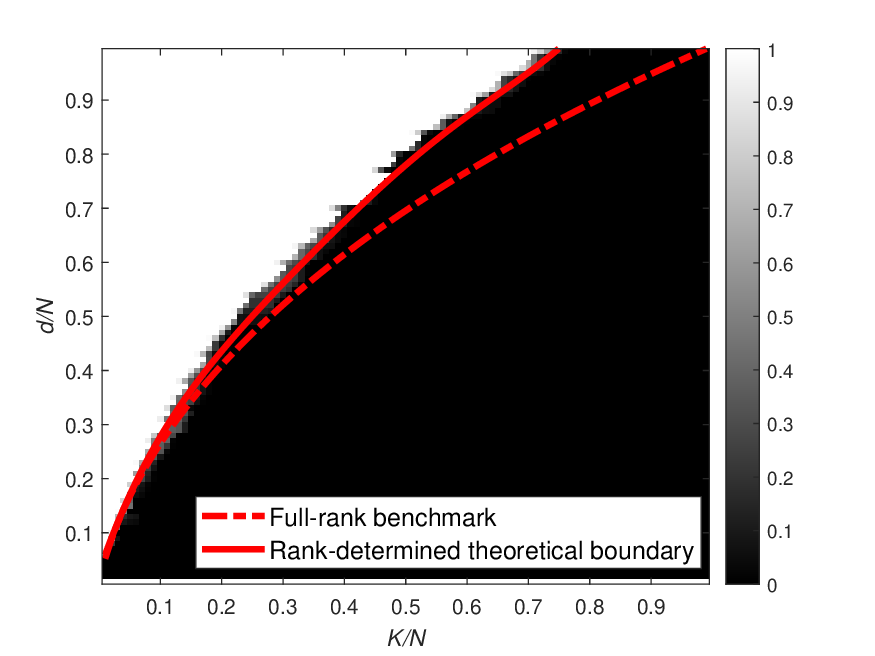}}
	\caption{Empirical phase transition of the {identifiability} of MLE under various signature matrices. $d:=\frac{L^2-L}{2}$ where $L$ is length of user signatures. $K$ is the number of active users and $N$ is the number of total users.}
	\label{fig:my_figure}
\end{figure*}

The following theorem is a direct consequence of Proposition \ref{Pro:Tropp}, together with explicit calculations of the asymptotic of $\delta(\mathcal{D})$. Its proof can be found in the appendix.

\begin{theorem}\label{The:main}
Denote $r :=  \text{rank}(\boldsymbol{A}_{\text{RI}})$. As $N, K, r \rightarrow \infty$ with fixed ratios $r/N \rightarrow \alpha \in  (0,1)$ and $K/N \rightarrow \epsilon \in (0,1)$, there exists a function $\delta_{*}(\epsilon)$ such that 
\begin{equation} \label{theorem_boundary}
    \begin{aligned}
        & \alpha > \delta_{*}(\epsilon) \Rightarrow  \mathrm{Pr} \left \{ \mathcal N({\boldsymbol{A}_{\text{semi-random}}}) \cap \mathcal{C} = \{ \boldsymbol{0}  \right \} \} \rightarrow 1, \\
        & \alpha < \delta_{*}(\epsilon) \Rightarrow  \mathrm{Pr} \left  \{ \mathcal N({\boldsymbol{A}_{\text{semi-random}}}) \cap \mathcal{C} \neq \{ \boldsymbol{0} \} \right \} \rightarrow 1,
    \end{aligned}
\end{equation}
where
\begin{equation}\label{boundary}
\delta_{*}(\epsilon):=  1 - (1-\epsilon) \Phi \left ( \mu_{*}(\epsilon) \right ),\quad\forall\epsilon\in(0,1),
\end{equation}
with $\mu_{*}(\epsilon)$ being the unique solution to the equation of $\mu$ in $(0,\infty)$:
\begin{equation}\label{Eqn:fixed_point_mu}
(1-\epsilon) \cdot \left( \mu \left (1-\Phi(\mu) \right) - \phi(\mu) \right) + \epsilon \mu = 0.
\end{equation}
Here, $\phi(\cdot)$ and $\Phi(\cdot)$ denote the PDF and CDF of the standard normal distribution. 

\end{theorem}

Theorem \ref{The:main} provides a precise characterization of the phase transition condition for the \textit{identifiability} of the MLE, under a semi-random model. We see that the phase transition boundary is determined by two factors: {(\romannumeral1) the statistical dimension of $\mathcal{D}$ (parameterized by $\epsilon$) and (\romannumeral2) the rank of $\boldsymbol{A}_{\text{RI}}$.}
We note that the discussions can be readily extended to the complex-valued cases. Details are omitted due to space limitation.

\subsection{Behavior of the Statistical Dimension in the Sparse Limit}

We analyze the behavior of the statistical dimension $\delta_{*}(\epsilon)$ as $\epsilon \rightarrow 0$. This provides a quantitative estimate of the sparsity-undersampling tradeoff of the MLE in the sparse limit.
\begin{proposition}\label{Pro:delta_limit}
Let $\delta_\ast(\epsilon)$ be defined as in \eqref{boundary}. We have
\begin{equation} \label{Asymptotic_result}
     \delta_{*}(\epsilon) \sim 2\epsilon\log \left( 1/\epsilon \right),\quad \epsilon\to0. 
\end{equation}
\end{proposition}
\begin{IEEEproof}
Recall that $\mu_\ast(\epsilon)\in(0,\infty)$ is defined as the unique solution to \eqref{Eqn:fixed_point_mu}, namely,
\begin{equation}\label{Eqn:fixed_point_mu2}
1 - \Phi\left(\mu_\ast(\epsilon)\right) - \frac{\phi\left(\mu_\ast(\epsilon)\right)}{\mu_\ast(\epsilon)} =-\frac{\epsilon}{1-\epsilon},\quad \forall \epsilon\in(0,1).
\end{equation}
Denote $G(\mu):=1-\Phi(\mu)-\frac{\phi(\mu)}{\mu}$. It is straightforward to verify that $G(\mu)$ is a monotonically increasing function on $\mu\in(0,\infty)$. Further, $\lim_{\mu\to\infty} G(\mu)=0$ and $\lim_{\mu\to0}G(\mu)=-\infty$. Therefore, $\mu_\ast(\epsilon)\to \infty$ as $\epsilon\rightarrow0$. Towards providing a quantitative estimate of $\mu_\ast(\mu)$, we first note that
\begin{equation}\label{Eqn:Phi_epsilon_mu2}
\begin{split}
\lim_{\epsilon \rightarrow 0}\frac{1-\Phi\left(\mu_\ast(\epsilon)\right)}{\epsilon \cdot \mu_\ast^2(\epsilon)}&\overset{(a)}{=}\lim_{\epsilon \rightarrow 0}\frac{1-\Phi\left(\mu_\ast(\epsilon)\right)}{-(1-\epsilon) \cdot G\left(\mu_\ast^2(\epsilon) \right) \cdot \mu_\ast^2(\epsilon)}\\
&\overset{(b)}{=}1,
\end{split}
\end{equation}
where step $(a)$ is from \eqref{Eqn:fixed_point_mu2}, step $(b)$ is due to $\Phi(-\mu)=1-\Phi(\mu)$ and the following elementary fact:
\[
\lim_{\mu\to \infty} \frac{1-\Phi(\mu)}{G(\mu)\cdot\mu^2} \overset{t=-\mu}{\Longrightarrow} \lim_{t \to -\infty} \frac{\Phi(t)}{\left(\Phi(t)+ \frac{\phi(t)}{t}\right)\cdot t^2}=-1.
\]
Taking the logarithm of \eqref{Eqn:fixed_point_mu2} and dividing both sides by $\mu^2_\ast(\epsilon)$ leads to ($\forall \epsilon\in(0,1)$)
\[
\frac{\log \phi(\mu_\ast(\epsilon))}{\mu_\ast^2(\epsilon)}=\frac{\log\left(\mu_\ast(\epsilon)\right)}{\mu_\ast^2(\epsilon)}+\frac{\log\left(1-\Phi\left(\mu_\ast(\epsilon)\right)+\frac{\epsilon}{1-\epsilon}\right)}{\mu_\ast^2(\epsilon)}.
\]
Sending $\epsilon\to0$ and noting that $\phi(\cdot)$ is the PDF function of the standard normal distribution, we obtain
\begin{equation}\label{Eqn:limit_final}
\frac{\log\left(1-\Phi\left(\mu_\ast(\epsilon)\right)+\frac{\epsilon}{1-\epsilon}\right)}{\mu^2_\ast(\epsilon)}\to-\frac{1}{2},\quad \epsilon\to0.
\end{equation}
Using \eqref{Eqn:Phi_epsilon_mu2}, we have
\begin{equation}\label{Eqn:limit_final2}
\frac{\log\left(1 - \Phi\left(\mu_\ast(\epsilon)\right)+\frac{\epsilon}{1-\epsilon}\right)}{\mu^2_\ast(\epsilon)}\sim\frac{\log\left(\epsilon\cdot \mu_\ast^2(\epsilon) \right)}{\mu^2_\ast(\epsilon)},\quad \epsilon\to0.
\end{equation}
Combining \eqref{Eqn:limit_final} and \eqref{Eqn:limit_final2} yields $\mu_\ast^2(\epsilon)\sim 2\log \epsilon^{-1} (\epsilon\to0)$. Using \eqref{Eqn:Phi_epsilon_mu2}, we have $1-\Phi\left(\mu_\ast(\epsilon)\right)\sim 2\epsilon\log\epsilon^{-1}$; thus[c.f. \eqref{boundary}]
\[
\delta_\ast(\epsilon):=1 - (1-\epsilon) \Phi \left ( \mu_{*}(\epsilon) \right )\sim 2\epsilon\log\epsilon^{-1},\quad \epsilon\to0.
\]
\end{IEEEproof}

The behavior of $\delta_\ast(\epsilon)$ in the sparse limit $\epsilon\to0$ agrees with existing results on $\ell_1$ minimization reconstruction for compressed sensing problems \cite{donoho2010precise}.

\section{Simulation Results} \label{Numerical_Results}

This section provides numerical examples to validate the theoretical results in Theorem \ref{The:main}. 
In our experiments, we used the LP approach in \cite{Phase_Yuwei_2022} to certify the feasibility problem in \eqref{convex_feasibility_event}. We conducted experiments under the following models of $\boldsymbol{S}$:
\begin{itemize}
	\item \textit{Gaussian}: the entries of $\boldsymbol S$ are IID standard Gaussian;
	\item \textit{Rademacher}: the entries of $\boldsymbol S$ are IID and take $\{+1,-1\}$ with equal probability;
	\item \textit{Doubly sub-sampled Hadamard}: $\boldsymbol S$ is a sub-sampled Hadamard matrix. Specially, the signature matrix is generated by randomly sub-sampling $L$ rows and $N$ columns from a $N_{\text{full}}\times N_{\text{full}}$ Hadamard matrix.
\end{itemize}
We set $N=5000$ in all experiments ($N_{\text{full}}=16384$ for the doubly sub-sampled Hadamard case). The heatmap represents the empirical probability of the event \eqref{convex_feasibility_event} calculated by 100 independent realizations.

Fig. 1 displays the empirical phase transition for \eqref{convex_feasibility_event} together with its theoretical predictions in Theorem \ref{The:main}. The theoretical curves displayed in Fig. 1 set the values of $\alpha$ in the following way. Note that $\alpha$ is the limit of $\alpha_N:=r/N$ as the problem sizes tend to infinity, where $r$ denotes the rank of $\bm{A}_{2}$; see \eqref{compact_null} for the definition of $\bm{A}_2$. For the Gaussian model and the Rademacher matrix, the matrices have full rank (i.e., $r=d$) with high probability (for the former case, with probability one). Hence, we set $\alpha=d/N$ for these two cases. For the doubly sub-sampled Hadamard model, however, the matrix $\bm{A}_{2}$ is rank deficient. Our numerical results suggest that $\alpha_N:=r/N$ converges and we set $\alpha$ to this limit. 
Fig. 1 shows that the theoretical results based on the semi-random model accurately describe the actual locations of the empirical phase transition for all three models.
Asymptotically, the Gaussian and Rademacher signatures outperform the Hadamard signatures.

\section{Conclusion} \label{Conclusion}

The phase transition analysis in this letter is based on a mathematically tractable semi-random model, which consists of an all-ones row and a rotationally-invariant random matrix. Numerical results suggest that theoretical predictions derived based on the semi-random model accurately describe the phase transition for the actual models with a complicated column-wise Kronecker product structure. Nevertheless, the validity of our analysis relies on the correctness of the universality conjecture. To this end, the recent spectral universality work \cite{Rishabh2024_Spectral_Universality} shed light on a possible fully rigorous proof.

\appendices
\section{Proof of Theorem \ref{The:main}}

Note that the statistical dimension of a linear subspace is exactly the subspace dimension itself. Theorem \ref{The:main} is a consequence of Proposition \ref{Pro:Tropp} together with the following lemma.

\begin{lemma}\label{Pro:stat-dim}
Let $\mathcal{D}$ be defined as in \eqref{Eqn:D_def}. As $N,K \rightarrow \infty$ with $K/N \rightarrow \epsilon \in (0,1)$, we have
\begin{equation} \label{normalized_statistical_dimension}
    \frac{1}{N}\delta(\mathcal{D}) \to \delta_{*}(\epsilon),
\end{equation}
where $\delta_{*}(\epsilon)$ is defined in \eqref{boundary}.
\end{lemma}

\begin{IEEEproof}
We first prove that the following holds:
\begin{subequations}\label{Eqn:stat-dim-opt-final}
\begin{equation}
\frac{1}{N}\|\Pi_{\mathcal{D}}(\bm{g})\|^2 = \min_{\mu\in\mathbb{R}}\ f_N(\mu;\bm{g}), \quad \forall \bm{g}\in\mathbb{R}^N,
\end{equation}
where $\Pi_{\mathcal{D}}(\cdot)$ is defined in \eqref{statistical_dimension_compute},
\begin{equation}\label{Eqn:f_N_def}
f_N(\mu;\bm{g}):=\frac{1}{N}\sum_{i\in\mathcal{I}} (g_i-\mu)_{+}^2 +\frac{1}{N}\sum_{i\in\mathcal{I}_c} (g_i-\mu)^2,
\end{equation}
\end{subequations}
and $\mathcal{I}:=\{1,\ldots,N-K\}$, $\mathcal{I}_c:=\{N-K+1,\ldots,N\}$. Towards proving \eqref{Eqn:stat-dim-opt-final}, we note
\begin{subequations}\label{optimization_problem}
\begin{eqnarray} 
\Pi_{\mathcal{D}}(\bm{g})= &    \underset{{\bm{x}_{\mathcal{I}}\ge \bm{0}_{N-K}}}{\text{argmin}}  \  &  \frac{1}{2}\Vert \boldsymbol{x} - \boldsymbol{g}\Vert^2 \\
  &  \text{s. t.}\   & \bm{1}^T\bm{x}=0.
\end{eqnarray}
\end{subequations}
We introduce a Lagrange multiplier for the equality constraint and consider the following Lagrangian:
\begin{equation}
    F(\boldsymbol{x},  \mu) = \frac{1}{2} \Vert \boldsymbol{x} - \boldsymbol{g}\Vert^2 + \mu \boldsymbol{1}^T \boldsymbol{g}.
\end{equation}
The minimum of $F(\bm{x},\mu)$ over the constraint set $\{\bm{x}\in\mathbb{R}^N \vline \bm{x}_{\mathcal{I}}\ge \bm{0}_{N-K}\}$ admits the following closed-form expression:
\begin{equation} \label{semi_solution}
    \begin{aligned}
        x_i^\star  &= (g_i - \mu)_{+}, \quad  i \in\mathcal{I}, \\ 
        x_i^\star &= (g_i - \mu), \quad  i\in\mathcal{I}_c,
    \end{aligned}
\end{equation}
where $(g_i - \mu)_{+} = \max(g_i - \mu, 0)$. The Lagrange multiplier $\mu\in\mathbb{R}$ is chosen such that $\bm{x}^\star$ satisfies the constraint $\bm{1}^T\bm{x}=0$:
\begin{equation}\label{Eqn:mu_equation}
\sum_{i\in\mathcal{I}} (g_i - \mu)_{+}+\sum_{i\in\mathcal{I}_c} (g_i - \mu) = 0.
\end{equation}
Note that the left-hand side (LHS) of \eqref{Eqn:mu_equation} is a strictly decreasing function of $\mu$ for any $\bm{g}\in\mathbb{R}^N$. Further checking the limit behaviors of it shows that \eqref{Eqn:mu_equation} has a unique solution, which we denote as $\mu_\ast$ in the sequel. To summarize, we have shown that $\frac{1}{N}\|\bm{x}^\star\|^2:=\frac{1}{N}\|\Pi_{\mathcal{D}}(\bm{g})\|^2$ satisfies the following
\begin{align*}
\frac{1}{N}\|\Pi_{\mathcal{D}}(\bm{g})\|^2 &=\frac{1}{N}\sum_{i\in\mathcal{I}} (g_i-\mu_\ast)_{+}^2 +\frac{1}{N}\sum_{i\in\mathcal{I}_c} (g_i-\mu_\ast)^2\\
&=f_N(\mu_\ast;\bm{g})\\
&= \min_{\mu\in\mathbb{R}}\ f_N(\mu;\bm{g}),
\end{align*}
where the last step is due to that $f_N(\mu;\bm{g})$ is a strongly convex function and its minimum is uniquely determined by its first order optimality condition, which is given by \eqref{Eqn:mu_equation}. This completes the proof of  \eqref{Eqn:stat-dim-opt-final}. 

The characterization in \eqref{Eqn:stat-dim-opt-final} is a deterministic result and holds for all $\bm{g}\in\mathbb{R}^N$. We now consider $\bm{g}\sim\mathcal{N}(\bm{0},\bm{I}_N)$ and prove the almost sure convergence of $\frac{1}{N}\|\Pi_{\mathcal{D}}(\bm{g})\|^2$ as $N,K\to\infty$ with $K/N\to\epsilon\in(0,1)$. By the strong law of large numbers, the following holds for any fixed $\mu\in\mathbb{R}$:
\begin{equation}
f_N(\mu;\bm{g})\overset{a.s.}{\longrightarrow} f(\mu),
\end{equation}
where
\begin{equation}\label{Eqn:f_def}
f(\mu):=(1-\epsilon)\mathbb{E}\left[(G-\mu)_{+}^2 \right]+\epsilon\mathbb{E}\left[(G-\mu)^2\right],
\end{equation}
and $G\sim\mathcal{N}(0,1)$. It is easy to see that $f(\mu)$ is a strongly convex function of $\mu$ and hence admits a unique minimum. By the convexity lemma \cite[Theorem 10.8]{rockafellar2015convex}, pointwise convergence of a sequence of convex functions implies uniform convergence over any compact set; see also \cite{pollard1991asymptotics}. The function $f(\mu)$ has a unique minimum. Using a standard ``artificial boundedness'' argument as in \cite[Lemma 10]{thrampoulidis2018precise}, we then conclude that
\begin{equation}
\min_{\mu\in\mathbb{R}}\ f_N(\mu;\bm{g})\overset{a.s.}{\longrightarrow} \min_{\mu\in\mathbb{R}}\ f(\mu).
\end{equation}
By \eqref{Eqn:stat-dim-opt-final} and the definition of $f(\cdot)$ in \eqref{Eqn:f_def}, we have
\begin{equation}\label{Eqn:stat-dim-as}
\frac{1}{N}\|\Pi_{\mathcal{D}}(\bm{g})\|^2\overset{a.s.}{\longrightarrow} \min_{\mu\in\mathbb{R}}\ (1-\epsilon)\mathbb{E}\left[(G-\mu)_{+}^2 \right]+\epsilon\mathbb{E}\left[(G-\mu)^2\right].
\end{equation}
Recall that $\frac{1}{N}\delta(\mathcal{D}):=\mathbb{E}\left[\frac{1}{N}\|\Pi_{\mathcal{D}}(\bm{g})\|^2\right]$. To translate \eqref{Eqn:stat-dim-as} into the convergence of $\delta(\mathcal{D})/N$, we next prove $\sup_N\mathbb{E}\left[\|\Pi_{\mathcal{D}}(\bm{g})\|^4/N^2\right]<\infty$, which guarantees the uniform integrability of $\frac{1}{N}\|\Pi_{\mathcal{D}}(\bm{g})\|^2$. By the polar decomposition \cite{Tropp}, $\|\bm{g}\|^2=\left\|\Pi_{\mathcal{D}}(\bm{g})\right\|^2+\left\|\Pi_{\mathcal{D}^o}(\bm{g})\right\|^2$, where $\mathcal{D}^o$ denotes the polar cone of $\mathcal{D}$. Hence, $\frac{1}{N}\left\|\Pi_{\mathcal{D}}(\bm{g})\right\|^2\le \frac{1}{N}\|\bm{g}\|^2$. When $\bm{g}\sim\mathcal{N}(\bm{0},\bm{I}_N)$, we have $\mathbb{E}[\|\bm{g}\|^4]=N^2+N$. Hence, $\mathbb{E}\left[\|\Pi_{\mathcal{D}}(\bm{g})\|^4/N^2\right]\le (N^2+N)/N^2\le 2$, $\forall N\ge1$.
This verifies the uniform integrability condition; hence
\begin{equation}\label{Eqn:stat-dim-expectation}
\frac{1}{N}\delta(\mathcal{D})\to \min_{\mu\in\mathbb{R}}\ (1-\epsilon)\mathbb{E}\left[(G-\mu)_{+}^2 \right]+\epsilon\mathbb{E}\left[(G-\mu)^2\right].
\end{equation}
It remains to check that the solution to the right-hand side (RHS) of \eqref{Eqn:stat-dim-expectation} is the unique solution to \eqref{Eqn:fixed_point_mu}. This follows from checking the first order optimality condition for \eqref{Eqn:fixed_point_mu}
$-(1-\epsilon)\mathbb{E}\left[(G-\mu)_{+}\right]+\mu\epsilon=0$, 
which recovers \eqref{Eqn:fixed_point_mu} together with $\mathbb{E}\left[(G-\mu)_{+}\right]=\phi(\mu)-\mu(1-\Phi(\mu))$, $\forall\mu\in\mathbb{R}$. It is straightforward to show that the unique solution is positive. This completes the proof.
\end{IEEEproof}

\balance
\bibliographystyle{IEEEtran}
\bibliography{IEEEabrv,ref}

\end{document}